# More on verification of probability forecasts for football outcomes: score decompositions, reliability and discrimination analyses.

Jean-Louis Foulley

IMAG-UMR 5149, Université de Montpellier 34090 Montpellier -France

**Abstract**

Forecast of football outcomes in terms of Home Win, Draw and Away Win relies largely on ex ante probability elicitation of these events and ex post verification of them via computation of probability scoring rules (Brier, Ranked Probability, Logarithmic, Zero-One scores). Usually, appraisal of the quality of forecasting procedures is restricted to reporting mean score values. The purpose of this article is to propose additional tools of verification, such as score decompositions into several components of special interest. Graphical and numerical diagnoses of reliability, resolution and discrimination and kindred statistical methods are presented using different techniques of binning (fixed thresholds, quantiles, logistic and iso regression). These procedures are illustrated on probability forecasts for the outcomes of 4 seasons of the UEFA Champions League (C1) at the end of the group stage based on typical Poisson regression models with reasonably good results in terms of reliability and resolution as compared to those obtained from bookmaker odds. Lack of forecasting draws efficiently due to poor resolution and discrimination is again outlined contrarily to home and away wins. Results are also discussed in the light of the Poisson regression model used and its features.

# 1 Introduction

The list of areas opening to forecast would be exceedingly long if one wishes to draw it up in details: from economic inflation, employment rates, weather and climate change, medical diagnosis and biological tests to media and entertainment market as well as gambling & sporting events among others. Forecasting consists first of producing forecasts from available data and methodologies, and second to assess their quality. Forecasts are basically of two types: pointwise or by means of probability. Quantify the uncertainty of a forthcoming event highlights the superiority of probability forecasts over categorical ones even economically (Winkler and Murphy 1979; Dawid 1986). As far as football matches are concerned, outcomes are in terms of either results such as Win, Draw or Loss (WDL)-also abbreviated as Home Win, Draw and Away Win (H,D,A) for two-legged matches- i.e., categorical data, or in terms of scorelines {Y(A), Y(B)} goals in match (A vs B) i.e., pairs of integers: see the review by Reade, Singleton, and Brown (2021). Here we will be concerned with probability forecasts of WDL (HDA) results. In this area, much effort has been devoted to statistical approaches to forecasting especially by modelling outcomes of matches (Scarf and Selliti-Rangel, 2019). There is a growing demand for relevant and acute probabilistic forecasting due to the large audience for TV broadcasting of association football matches and the related huge betting markets. Traditionally in football, one summarizes HDA forecast performance by a synthetic criterium such as the mean square error known as Brier's score (1950). Although this statistic brought useful information, it is only an average measure of the overall accuracy of such predictions. Analytical attributes of this accuracy can be brought out such as Reliability, Resolution, Discrimination, Refinement to shade additional light on some qualities or deficiencies of issued forecasts. The purpose of this note is mostly pedagogical with the aim to review the main tools available to that respect, and to illustrate them on data of a well-known club competition with the UEFA champions League (also known in brief as C1) forecasted by simple Poisson regression models. Focus is on a distribution-oriented approach based on the joint distribution of elementary forecast, binary outcome pairs and its factorization. We will deal with several decompositions of the Brier score (BRS) applied to each binary outcome considered separately namely 1) the Murphy decomposition derived from the so-called Calibration-Refinement factorization (CR), and 2) the Likelihood-base (LB) and Yates's decompositions. In parallel to the CR decomposition, we will present a graphical device known as the reliability diagram which allows diagnostic of strength or deficiency in this component. In each case, comparison will be made in section 3 between forecasts derived from a simple Poisson loglinear model (POI), and to Bookmaker Odds implied probabilities (ODD). Finally, in the last ($4^{th}$) section, we discuss the main points raised by using these decompositions and possible implications for improving forecast efficiency.

# 2 Decomposition of the Brier score

## 2.1 Basic theory

Let $X$ be the binary outcome of the event H with probability $q$ and $P$ the random variable probabilistic forecast of $X$ taking values $p$. Taking as scoring rule, the quadratic or Half-Brier Score defined as the loss function $S(P,X)=(P-X)^2$, and using the conditioning de conditioning rule, the Murphy (1973) decomposition of its expectation can be written as

$$\mathbb{E}[S(P,X)]=\text{Var}(X)-\mathbb{E}_P\left\{\left[\mathbb{E}_X(X\mid P)-\mathbb{E}_X(X)\right]^2\right\}+\mathbb{E}_P\left\{\left[\mathbb{E}_X(X\mid P)-P\right]^2\right\}, \quad (1)$$

where expectation is taken with respect to the pairs of forecasts and outcomes $P,X$.

Examination of this formula immediately identifies the 3 components of this decomposition:

1) Uncertainty (UNC) equal to $\text{Var}(X)=q(1-q)$, the variance of the outcome that is out of control of the forecaster,
2) Resolution (RES) equal to $\mathbb{E}_P\left\{\left[\mathbb{E}_X(X\mid P)-\mathbb{E}_X(X)\right]^2\right\}$ referring to the variability between the conditional expectations of the observed outcomes given their forecasts around the mean outcome,
3) Reliability (REL) or Calibration equal to $\mathbb{E}_P\left\{\left[\mathbb{E}_X(X\mid P)-P\right]^2\right\}$ measuring the average squared differences between the conditional expectation of the outcome and its forecast.

Murphy and Winkler (1987) also gave the dual decomposition of (1)

$$\mathbb{E}[S(P,X)] = \mathrm{Var}(P) - \mathrm{Var}_X\left[\mathbb{E}_P(P \mid X)\right] + \mathbb{E}_X\left\{\left[\mathbb{E}_P(P \mid X) - X\right]^2\right\}. \qquad (2)$$

This decomposition is known as the Likelihood-base factorization as opposed to the Calibration-Refinement factorization of formula (1) referring to the 2-way decomposition of the joint distribution of $(P,X)$. This formula leads to identify three components: i) Refinement (REF) equal to $\mathrm{Var}(P)$, the variance of probabilistic forecasts also known as Sharpness, ii) Discrimination (DIS) equal to $\mathrm{Var}_X\left[\mathbb{E}_P(P \mid X)\right]$ i.e., the variance between conditional distributions of forecasts given the outcomes $X$, iii) Conditional bias type 2 as called by Bradley et al. (2003) and equal to $\mathbb{E}_X\left\{\left[\mathbb{E}_P(P \mid X) - X\right]^2\right\}$ which is the counterpart of resolution. A special case of interest is the one with $P$ having a probability mass function (pmf) concentrated at $p = \mathbb{E}(X)$, the mean of the marginal distribution of the binary outcome (the so-called climatological forecast). Then the forecast has no refinement $(\mathrm{Var}\,P = 0)$, no discrimination as well, necessarily no resolution, even though it is perfectly calibrated. In fact, it is the only forecast being both reliable with no discrimination (Bröcker, 2012). In that case, $\mathbb{E}[S(P,X)]$ reduces to its uncertain component UNC as an upper reference value for the expected Brier score. That is the reason why the decomposition in (1) is often expressed as fractions of UNC and the complement to one of the scaled Brier Score (BS) as a Brier Skill Score

$$BSS = 1 - BS / BS_{ref} = (REL\text{-}RES) / UNC. \qquad (3)$$

This formula clearly emphasizes the trade-off between these two components with the aim of increasing resolution without neglecting reliability.

There is another decomposition by Yates (1982) deserving as much attention as it singularizes and enjoys influential components of the expected quadratic score which are easily interpretable.

$$\begin{aligned}\mathbb{E}[S(P,X)] = {} & \mathrm{Var}(X) - 2\mathrm{Cov}(P,X) + \mathrm{Var}_X\left[\mathbb{E}_P(P \mid X)\right] \\ & + \mathbb{E}_X\left[\mathrm{Var}_P(P \mid X)\right] + \left[\mathbb{E}(P) - \mathbb{E}(X)\right]^2\end{aligned}. \qquad (4)$$

This formula stems from the basic expression $\mathbb{E}[S(P,X)] = \mathrm{Var}(P-X) + \left[\mathbb{E}(P) - \mathbb{E}(X)\right]^2$ with $\mathrm{Var}(P) = \mathrm{Var}_X\left[\mathbb{E}_P(P \mid X)\right] + \mathbb{E}_X\left[\mathrm{Var}_P(P \mid X)\right]$ and $\mathbb{E}(P \mid X)$ being the best predictor of $P$ given $X$. As this predictor is a linear one: $\mathbb{E}(P \mid X) = a + bX$ with $a = \mathbb{E}(P \mid X = 0)$, $b = \mathbb{E}(P \mid X = 1) - \mathbb{E}(P \mid X = 0)$, then $\mathrm{Var}_X\left[\mathbb{E}_P(P \mid X)\right] = b^2 \,\mathrm{Var}(X)$, and $\mathrm{Cov}(P,X) = b\,\mathrm{Var}(X)$. These last two expressions highlight the key role of the regression coefficient $b$ of $P$ on $X$ equal to the difference in expected value of forecasts pertaining to future positive outcomes from those out of negative ones.

Yates (1982) emphasized the different influence of the two components of $\mathrm{Var}(P)$. The first one, which is the between class variance $\mathrm{Var}_X\left[\mathbb{E}_P(P \mid X)\right]$, he qualified as "VarPmin", is beneficial. The second one, which is the within class variance $\mathbb{E}_X\left[\mathrm{Var}_P(P \mid X)\right]$, called "ΔVarP" or "Scattered Variance" representing the lack of sharpness of the distributions of $P \mid X = x$, is detrimental. The last term in (4) measures marginal bias and was called "Calibration or Reliability-in-Large" (CIL or RIL). In short, this decomposition is written as: UNC- 2COV+ VarPmin+ ΔVarP+ RIL. Yates' and the LB decompositions are closely related. Actually, VarPmin coincides exactly with DIS, and the sum VarPmin+ΔVarP equals REF, and CB2 is UNC-2COV+VarPmin+RIL.

Table 1: Calibration-Refinement decomposition of Brier's score pertaining to Home Win (A), Draw (B) and Away Win (C) under two forecasting procedures: Poisson regression model (POI) and Odds (ODD)

| A-HWIN | BRS | SKI (%) | B-TEST | MET | REL | RES |
|---|---|---|---|---|---|---|
| | | | | INT | 0.0035 (1.4) | 0.0644 (26.2) |
| POI | 0.1849 | 24.8 | 1.035 | QUA | 0.0030 (1.2) | 0.0639 (26.0) |
| | | | [0.309] | ISO | **0.0116 (4.7)** | **0.0725 (29.5)** |
| | | | | INT | 0.0041 (1.7) | 0.0766 (31.2) |
| ODD | 0.1732 | 29.5 | 2.099 | QUA | 0.0048 (2.0) | 0.0774 (31.5) |
| | | | [0.147] | ISO | **0.0122 (4.9)** | **0.0847 (34.5)** |

UNC=0.2458

| B-DRAW | BRS | SKI (%) | B-TEST | MET | REL | RES |
|---|---|---|---|---|---|---|
| | | | | INT | 0.0010 (0.5) | 0.0036 (1.9) |
| POI | 0.1849 | 1.4 | 3.995 | QUA | 0.0031 (1.7) | 0.0058 (3.1) |
| | | | [0.045] | ISO | **0.0099 (5.3)** | **0.0125 (6.7)** |
| | | | | INT | 0.0011 (0.6) | 0.0066 (3.5) |
| ODD | 0.1820 | 3.0 | 2.102 | QUA | 0.0005 (0.3) | 0.0060 (3.2) |
| | | | [0.147] | ISO | **0.0048 (2.6)** | **0.0092 (4.9)** |

UNC=0.1875

| C-AWIN | BRS | SKI (%) | B-TEST | MET | REL | RES |
|---|---|---|---|---|---|---|
| | | | | INT | 0.0047 (2.2) | 0.0505 (23.4) |
| POI | 0.1700 | 21.2 | 0.001 | QUA | 0.0029 (1.3) | 0.0487 (22.5) |
| | | | [0.975] | ISO | **0.0078 (3.6)** | **0.0537 (24.9)** |
| | | | | INT | 0.0027 (1.3) | 0.0616 (28.5) |
| ODD | 0.1569 | 27.3 | 0.582 | QUA | 0.0045 (2.1) | 0.0635 (29.4) |
| | | | [0.445] | ISO | **0.0100 (4.6)** | **0.0689 (31.9)** |

UNC=0.2158

BRS=REL-RES+UNC with REL: Reliability, RES: Resolution, UNC: Uncertainty) according to forecasting procedures (Poisson model (POI) vs Odds implied probabilities (ODD)), and different binning procedures (INT: Interval; QUA: Quantile: ISO-Regression).and expressed both in absolute value and p.100 of uncertainty (UNC) .
Skill (SKI) defined as SKI= (BRSref -BRS)/ BRSref where BRSref=UNC so that SKI=(RES-REL)/UNC.
B-TEST: Brier-Score Test for departure of its expectation from that induced by the null hypothesis of perfect forecast calibration expressed with its corresponding statistic and P-value within brackets.

## 2.2 Estimation from data

Practically, verification takes place from a data sample made of pairs $\{(p_i, x_i), i = 1,...,N\}$ of ex ante probabilistic forecasts $p_i$ and ex post binary outcomes $x_i$. Most quantities introduced previously can be estimated by their regular moment estimators. The expected quadratic score $\mathbb{E}[S(P, X)]$ is traditionally estimated by the empirical score:

$$\bar{S}(\mathbf{p}) = N^{-1}\sum_{i=1}^{N} S\left(p_i, x_i\right). \quad (5)$$

For the CR decomposition, REL and RES require estimations of $\mathbb{E}\left(X \mid P\right)$. If the forecasts take $K$ distinct values $\{p_k, k = 1,...,K\}$ with $n_k$ occurrences of binary outcomes $X$, then $\hat{\mathbb{E}}\left(X \mid P = p_k\right) = \bar{X}_k = X_{k+} / n_k$ with $X_{k+} = \sum_{i=1}^{N} I\left(p_i = p_k\right) X_i$ and $\bar{X} = \left(\sum_{i=1}^{N} X_i\right)/N$. In such cases, the Murphy (1973) decomposition is fully applicable without restrictions:

$$REL = N^{-1}\sum_{k=1}^{K} n_k \left(\bar{x}_k - p_k\right)^2,\ RES = N^{-1}\sum_{k=1}^{K} n_k \left(\bar{x}_k - \bar{x}\right)^2,\ UNC = \bar{x}\left(1 - \bar{x}\right). \quad (6)$$

In fact, in many applications as in forecasting Football match results, we stay in-between discrete and continuous cases, facing many distinct forecast values. In such cases, forecasts have to be distributed into intervals named bins $B_1,..B_d,..,B_D$ and averaged within bins i.e., letting $I_d = \{i : p_i \in B_d\}$, pairs $\{(p_d, x_d), d = 1,...,D\}$ are computed as $p_d = n_d^{-1}\sum_{i\in I_d} p_i$, $\hat{x}_d = x_{d+} / n_d$ where $n_d = \# I_d$, $\hat{x}_{d+} = n_d^{-1}\sum_{i\in I_d} x_i$. To avoid inconsistencies in the CR decomposition, two extra components of within bin variance and covariance must be added to those in (6). We skip such complications by adopting a simple procedure as advocated by Siegert (2017).

Letting as in (5) $\bar{S}(\hat{\mathbf{x}}) = N^{-1}\sum_{i=1}^{N} S\left(f_i, \hat{x}_i\right)$ and $\bar{S}(\bar{x}\mathbf{1}_N) = N^{-1}\sum_{i=1}^{N} S\left(f_i, \bar{x}\right)$, the components of the mean score $\bar{S}(\mathbf{p})$ reduce to $REL = \bar{S}(\mathbf{p}) - \bar{S}(\hat{\mathbf{x}})$, $RES = \bar{S}(\bar{x}\mathbf{1}_N) - \bar{S}(\hat{\mathbf{x}})$, $UNC = \bar{S}(\bar{x}\mathbf{1}_N)$. (7)

This decomposition automatically satisfies the equality $\bar{S}(\mathbf{p}) = REL - RES + UNC$, and is equivalent to the original Murphy decomposition in the case of distinct discrete forecasts. It also ensures that i) resolution is nil when $\mathbf{p}$ is perfectly calibrated ($\mathbf{p} = \hat{\mathbf{x}}$), and ii) the constant climatological forecast $\mathbf{p} = \bar{x}\mathbf{1}_N$ is the only forecast satisfying RES=REL=0. Finally, it is potentially applicable to other proper probabilistic scoring rules, as the ignorance score $L\left(P, X\right) = -X\log(P) - (1 - X)\log(1 - P)$. Moreover, the statistic $2N\left[\bar{L}(\mathbf{p}) - \bar{L}(\hat{q}_p)\right]$ is an analog of the log-likelihood ratio statistic for a perfectly reliable forecast as an asymptotic Chi-square distribution with degrees of freedom equal to the number of parameters specifying the model for $q_p = \Pr(X = 1 \mid P = p)$. Different binning techniques are available such as fixed threshold intervals and fixed quantile intervals with potential optimization of their number (Bröcker, 2012; Gweon and Yu, 2019). A promising one relies on the non-parametric isotonic regression implemented via the pool-adjacent-violators (PAV) algorithm with optimality properties (Dimitriadis, Gneiting, and Jordan 2021).

It also provides a reliability diagram featuring graphically the CR decomposition of the joint distribution $[P, X] = [P][X \mid P]$ of binned data by the marginal (refinement) distribution of forecasts and plots of (re) calibrated probabilities $\hat{x}_d$ against automatically binned forecasts $p_d$.

Table 2: Calibration analysis via fitting a logistic model of the probability of Home win, Draw and Away win on the logit of its probabilistic forecast under a Poisson regression model (A: POI) and Odds implied probabilities (B: ODD)

A) POI

| Category | Criterion | Estimation | SE | T-Statistics | DF | P-value |
|---|---|---|---|---|---|---|
| Home Win | intercept | -0.259 | 0.119 | 4.700 | 1 | 0.030 |
| | slope | 1.113 | 0.129 | 0.765 | 1 | 0.382 |
| | D0 vs D1 | 423.085 vs 417.489 | | 5.596 | 2 | 0.061 |
| Draw | intercept | 0.153 | 0.466 | 0.108 | 1 | 0.742 |
| | slope | 0.932 | 0.346 | 0.039 | 1 | 0.843 |
| | D0 vs D1 | 426.981 vs 422.981 | | 4.000 | 2 | 0.135 |
| Away Win | intercept | 0.076 | 0.149 | 0.261 | 1 | 0.610 |
| | slope | 1.053 | 0.134 | 0.156 | 1 | 0.693 |
| | D0 vs D1 | 389.458 vs 389.176 | | 0.282 | 2 | 0.870 |

B) ODD

| Category | Criterion | Estimation | SE | T-Statistics | DF | P-value |
|---|---|---|---|---|---|---|
| Home Win | intercept | -0.195 | 0.123 | 2.493 | 1 | 0.114 |
| | slope | 1.175 | 0.129 | 1.843 | 1 | 0.174 |
| | D0 vs D1 | 400.472 vs 395.864 | | 4.608 | 2 | 0.100 |
| Draw | intercept | 0.399 | 0.416 | 0.350 | 1 | 0.454 |
| | slope | 1.170 | 0.325 | 0.021 | 1 | 0.601 |
| | D0 vs D1 | 418.832 vs 416.103 | | 2.729 | 2 | 0.255 |
| Away Win | intercept | 0.096 | 0.149 | 0.261 | 1 | 0.522 |
| | slope | 1.135 | 0.131 | 1.060 | 1 | 0.303 |
| | D0 vs D1 | 365.652 vs 364.515 | | 1.137 | 2 | 0.566 |

Intercept ($\alpha$) and slope ($\beta$) with their estimation and standard error (SE)

Deviance D(k)=-2L(k) where L(k) is the loglikelihood of the null model ($\alpha=0; \beta=1$) vs the unspecified parameter model

T-statistics: Wald for intercept=0 and slope=1; Deviance differences $\Delta D = D0 - D1$ and their corresponding degrees of freedom (DF) and P-values

To that respect, another way to assess Reliability via the conditional distribution $[X \mid P]$ of outcomes X given P is through a regression model, but in the framework of logistic instead of linear regression chosen by Reade, Singleton, and Brown (2021). Following Cox (1958), the model relating X to P is written via a logit linear predictor: $\text{logit}\left[\Pr\left(X_i = 1\right)\right] = \alpha + \beta \text{logit}\left(p_i\right)$ with $\alpha = 0$ and $\beta = 1$ for perfect reliability and typical patterns of reliability diagrams with i) $(\alpha > 0, \beta = 1)$ for concave under-forecasting profiles, ii) $(\alpha < 0, \beta = 1)$ for convex over-forecasting profiles as well as iii) $(\alpha = 0, \beta > 1)$ for sigmoid, and iv) $(\alpha = 0, \beta < 1)$ for inverse-sigmoid profiles. Statistical tests are available (Wald and likelihood ratio tests) for challenging the different hypotheses about such patterns.

# 3 Application

The purpose of this illustration is to assess the performance of probability forecasts of outcomes of the UEFA champions league (the so called C1) matches played during the group stage (GS). Four seasons were considered from 2017 to 2020. Forecast is based on a simple log linear Poisson regression model applied to score lines with intercept, home effect and two time-dependent ELO team covariates (see appendix A for a detailed description). The models were fitted to ex ante data, namely score lines of all the matches played during the 3 previous seasons e.g., 2017, 2018 and 2019 as training sample to forecasts of the 96 GS matches of 2020; the same applies to forecasts of the 2019 GS based on 2016, 2017 and 2018 seasons and so on. Inference about parameters of the Poisson loglinear model is based on posterior distributions and probability forecasts are obtained as expectations of predictive distributions. Computations are carried out via the Win/OpenBUGS software. As top reference, we considered the classical Bookmaker Odds (ODD) as 3-Way Odds implied Probabilities with probabilities derived as $p_{m,j} = o_{m,j}^{-1} / \sum_{k=1}^{3} o_{m,k}^{-1}$ where $o_{m,j}$ is the betting odd for $j=1,2,3$ (WDL) edited by OddsPortal (here an average of 10 to 12 odds from well-known betting companies).

Table 3: Characteristics of conditional distributions of probability forecasts given the outcomes under two Forecasting procedures: Poisson regression (POI) and Odds Probabilities (ODD)

| Method | | Home Win | | Draw | | Away Win | |
|---|---|---|---|---|---|---|---|
| | | POI | ODD | POI | ODD | POI | ODD |
| Sample sizes | | 217-167 | | 288-96 | | 263-121 | |
| Mean % | X=0 | 37.88 | 35.01 | 20.22 | 20.97 | 24.30 | 23.24 |
| | X=1 | 63.08 | 62.73 | 22.34 | 23.89 | 44.84 | 48.62 |
| | Dif 1-0 | 24.20 | 27.71 | 2.02 | 2.93 | 20.54 | 25.38 |
| Wilcoxon | Z | 9.93 | 10.76 | 3.59 | 3.55 | 9.09 | 10.13 |
| | P-val | *<0.0001* | *<0.0001* | *0.0002* | *0.0002* | *<0.0001* | *<0.0001* |
| KS | D | *0.473* | *0.511* | *0.236* | *0.236* | *0.447* | *0.521* |
| | P-val | *<0.0001* | *<0.0001* | *0.0007* | *0.0007* | *<0.0001* | *<0.0001* |
| C-statistic | Estimation | 0.795 | 0.820 | 0.622 | 0.624 | 0.789 | 0.820 |

Sample sizes of forecasts having X=0 vs X=1 respectively; Z: Normal approximation of the Wilcoxon-test with one sided P value; KS: Kolmogorov-Smirnoff two sample test on Max [F(X=0)-F(X=1)] C-statistic: Harrell's concordance index varying from 0.5 (no discrimination) to 1 (perfect discrimination) equal to AUC (area under the ROC curve)

## 3.1 CR decomposition of Brier's score

Due to the large number of unique probability profiles (377 among N=384 matches), forecasts were binned in three different ways: i) Fixed threshold intervals: D=10 from 0.0 to 1.0 for Home Win; D=5 with bounds at 0.10,0.15,0.20,0.25 and 0.35 for Draw and D=8 for Away Win with the first 7 bins equally spaced from 0 to 0.7 and the last one from 0.7; ii) Quantile thresholds intervals: deciles for Home Win and Away Win and quintiles for Draws iii) Bins automatically determined by the pool-adjacent-violators (PAV) algorithm used to set up the nonparametric isotonic regression deployed to estimate the conditional $q_p = \Pr(X=1 \mid P=p)$ outcome probabilities by minimizing the regression MSE with respect to D:

$$\sum_{d=1}^{D}\sum_{i=1}^{N} I\left(p_i \in \left[b_d, b_{d+1}\right]\right)\left(q_d - p_i\right)^2$$

under the constraints of isotonicity ($q_d$ estimation is a non-decreasing function of the original $p_i$'s). Results are displayed in Table 1, for Home Win, Draw and Away Win categories considered separately in terms of absolute values of Mean Brier score and its components (REL, RES, ACC) and Skill. Lack of reliability turns out to be small (lower than 5.5% of UNC) with miscalibration estimated a little bit higher under iso-regression, but satisfactory. These values are supported by statistics and P values of Brier's score tests of departure from zero miscalibration (Spiegelhalter, 1986; Sellier-Moiseiwitch and David, 1993).

Skill values are also appreciable for Home Win (25 to 30%) and Away Win (20 to 30%) with a little advantage of Odds vs Poisson of around 5%: 29.5 vs 24.8% for Home Win and 27.3 vs 21.2% for Away Win. On the opposite, skill remains quite poor for Draw: 1.4 to 3.0% for Poisson and Odds respectively. Corresponding reliability diagrams were produced with an example shown for Home Win under Poisson and Odds forecasts on Fig 1. There is some evidence of over-forecasting for Home Win for Poisson and Odds especially below p<0.60 with correlated under-forecasted Draws but calibrated values remain within the 95% pointwise consistency bars. These conclusions are confirmed by the results (table 2) of the logistic regression with intercept and slope calibration coefficients showing clearly over-forecasting of home wins $(\hat{\alpha}=-0.26, \hat{\beta}=1.11)$ with P values=0.03 and 0.38 respectively while Away wins are very well calibrated $(\hat{\alpha}=0.076, \hat{\beta}=1.053)$ .

*Figure 1: Reliability Diagrams for HomeWin Probability Forecasts with plots of the Conditional Probability Events (CEP) against the Forecast Probability Values via Iso Regression.*

## 3.2 LB and Yates' decompositions of Brier score

Likelihood-base as well as Yates' decomposition of Brier Score rely largely on the concept of discrimination between the conditional distributions of forecasts with positive outcomes vs forecast with negative outcomes. Characteristics of these two distributions are given in Table 3 and Fig 2. Differences between the means of these two conditional distributions are much more marked for Home Win and Away Win than for Draw. Again, these differences are more strongly marked with Odds than Poisson, as also reported by the Harrell c-statistics around 0.8 for Home Win and Away Win, and only 0.6 for Draw. Graphically, the boxplots in figure 2 confirmed this situation showing a clear separation of the two conditional distributions for Home Win and Away Win and a tiny one for Draw.

Table 4: Yates's and LB decompositions of Brier's score pertaining to Home Win, Draw and Away Win under two forecasting procedures: Poisson regression model (POI) and Odds probabilities (ODD)

| **POI** | Home Win | | Draw | | Away Win | | ALL | |
|---|---|---|---|---|---|---|---|---|
| | value | % | value | % | value | % | value | % |
| UNC | 0.2458 | 100.0 | 0.1875 | 100.0 | 0.2158 | 100.0 | 0.6490 | 100.0 |
| (-2) COV | **-0.1190** | **-48.4** | **-0.0076** | **-4.0** | **-0.0886** | **-41.1** | **-0.2150** | **-29.5** |
| VPB | 0.0144 | 5.8 | 0.0001 | 0.0 | 0.0091 | 4.2 | 0.0236 | 3.6 |
| VPW | 0.0413 | 16.8 | 0.0031 | 1.7 | 0.0336 | 15.6 | 0.0780 | 12.0 |
| RIL | 0.0024 | 1.0 | 0.0017 | 0.9 | 0.0000 | 0.0 | 0.0042 | 0.6 |
| REF | 0.0556 | 22.6 | 0.0032 | 1.7 | 0.0427 | 19.8 | 0.1016 | 15.6 |
| -DIS | -0.0144 | -5.8 | -0.0001 | -0.0 | -0.0091 | -4.2 | -0.0236 | -3.6 |
| CB2 | 0.1436 | 58.4 | 0.1817 | 96.9 | 0.1363 | 63.2 | 0.5396 | 83. |
| BRS | 0.1849 | 75.2 | 0.1848 | 98.6 | 0.1700 | 78.8 | 0.5397 | 83.1 |

| **ODD** | Home Win | | Draw | | Away Win | | ALL | |
|---|---|---|---|---|---|---|---|---|
| | value | % | value | % | value | % | value | % |
| UNC | 0.2458 | 100.0 | 0.1875 | 100.0 | 0.2158 | 100.0 | 0.6490 | 100.0 |
| (-2) COV | **-0.1362** | **-55.4** | **-0.0110** | **-5.9** | **-0.1095** | **-50.7** | **-0.2567** | **-39.6** |
| VPB | 0.0189 | 7.7 | 0.0002 | 0.1 | 0.0139 | 6.4 | 0.0329 | 5.1 |
| VPW | 0.0435 | 17.7 | 0.0042 | 2.2 | 0.0367 | 17.0 | 0.0844 | 13.0 |
| RIL | 0.0013 | 0.5 | 0.0011 | 0.6 | 0.0000 | 0.0 | 0.000 | 0.4 |
| REF | 0.0624 | 25.4 | 0.0044 | 2.3 | 0.0506 | 23.4 | 0.1173 | 18.1 |
| -DIS | -0.0189 | -7.7 | -0.0002 | -0.1 | -0.0139 | -6.4 | -0.0329 | -5.1 |
| CB2 | 0.1297 | 52.8 | 0.1778 | 94.8 | 0.1202 | 55.7 | 0.4276 | 65.9 |
| BRS | 0.1732 | 70.5 | 0.1820 | 97.1 | 0.1569 | 72.7 | 0.5120 | 78.9 |

Yates's decomposition into 5 components as follows BRS=UNC-2COV+VPB+VPW+RIL with UNC: Uncertainty, COV: Covariance between forecast and outcome, VPB: Variance among means of probability forecasts with outcome=1 and outcome=0, VPW: Average of Within groups variance and RIL marginal bias squared between the two groups, according to forecasting procedures (POI and ODD models).
Likelihood base decomposition into 3 components such that BRS=REF-DIS+CB2 with REF: Refinement or Sharpness of forecast variance, DIS=Discrimination same as VPB and CB2: Type 2 bias equal to VPW-2COV+RIL.

Detailed accounts of Yates' and LB decompositions are shown on Table 4. In short, what emerges from them lies in the large role and weight given to the covariance component: 41 to 48% for Home Win and Away Win under Poisson and 50 to 55% with Odds with nevertheless a non-negligible part devoted to "noise" variance VPW of forecasts (15 to 18%). The same picture applies to Draw but with much more tiny components, especially discrimination and covariance.

# 4 Discussion

This presentation was deliberately restricted to the most popular (strictly) proper scoring rules as this properness property is a cornerstone of decision theory based on minimizing expected loss (or maximizing utility) (Bernardo 1979, Gneiting and Raftery 2007). They provide an incentive for ex ante honesty and reward ex post accuracy. Little was said about verifying probability forecasts for multiple categories taken simultaneously. Probability scoring rules are extended easily to that situation as shown in table 3 for Yates' decomposition. Unhappily, such an extension is not straightforward for the CR decomposition. One reason for that lies on how to define bins for multiple categories. Procedures have been proposed to that respect by Bröcker (2012) based on some functions of the probability vector for the J multiple categories of interest. A simple way to handle the multiclass setting is by treating the problem as J one-versus-all binary events via e.g., a logistic-type regression with standard normalization of outcome probabilities. More generally, such forecasting verification methods already gained much attention in other fields especially in machine learning especially due to miscalibration of neural networks and its applications to health and medicine (Guo, Pleiss, and Weinberger 2017).

*Here Figure 2: Boxplots of the conditional distributions of probabilistic forecasts of binary events (Home Win, Draw and Away Win) given the observed outcomes (X=0) and (X=1) and according to the Poisson loglinear regression (poi)*

Although much of the theory and applications originated from meteorological literature (Winkler and Murphy 1979, Jolliffe and Stephenson 2003), there had been a few attempts to apply some of these analytical procedures to football match results, especially in the EPL (Selliti Rangel 2018, Wheatcroft 2019; Reade , Singleton, and Brown 2021) but not enough. This area would benefit from a more systematic utilization.

The Poisson Regression model serving here as a basis for forecasting probabilities of outcomes is a simple one in its category reduced here to a parsimonious version with only four parameters. Several more sophisticated variants of it have been reported in the literature (Dixon and Coles 1997, Karlis and Ntzoufras 2008, Groll et al. 2018; Dagaev and Rudyak 2019).

Table 5: Elementary decomposition of forecasting ability of Poisson regression (POI) and Odds Probabilities (ODD) for Home Win, Draw and Away Win in the UEFA 2017-2020 Champions League

| | Home Win | | Draw | | Away Win | |
|---|---|---|---|---|---|---|
| | POI | ODD | POI | ODD | POI | ODD |
| Brier score* | 0.185 | 0.173 | 0.185 | 0.182 | 0.170 | 0.157 |
| Skill% | 24.8 | 29.5 | 1.4 | 3.0 | 21.2 | 27.3 |
| Reliability* % | 4.7 | 4.9 | 5.3 | 2.6 | 3.6 | 4.6 |
| Resolution % | 29.5 | 34.5 | 6.7 | 4.9 | 24.9 | 31.9 |
| Discrimination % | 5.8 | 7.7 | 0.0 | 0.1 | 4.2 | 6.4 |
| ΔVariance P* % | 16.8 | 17.7 | 1.7 | 2.2 | 15.6 | 17.0 |
| 2Cov (P, X) % | 48.4 | 55.4 | 4.0 | 5.9 | 41.0 | 50.7 |

*The smaller, the better; % of Uncertainty
BRS: (Half) Brier score=REL-RES+UNC with REL: Reliability, RES: Resolution, UNC: Uncertainty
SKI= (BRSref -BRS)/ BRSref where BRSref=UNC so that SKI=(RES-REL)/UNC
Discrimination: Variance among means of probability forecasts with outcome=1 and 0, same as VarP min
ΔVariance P=VarP-VarP min i.e., average of within groups variance
COV (P, X): Covariance between forecast and outcome.

Despite is simplicity, this model offers several valuable features. First as pointed out by Pinheiro (2002) "Incorporating covariates in a non-linear mixed model (here generalized linear model) allows intergroup variation (here teams abilities) to be explained via fixed effects" and "the inclusion of covariates in the model leads to the removal of random effects". Second, the covariates for team abilities are represented by ELO ratings of teams, the advantages of which (free model dynamic rating, more weight to recent matches, time varying strengths) have been persuasively outlined by Hvattum and Arntzen (2010) and Aldous (2017). These ratings are especially welcomed at the start of the season (group stage) with none or very few matches already being played. Another benefit from this model lies in its property to automatically generate a negative correlation between Home and Away goals ( $\rho_{12}^{(M)} = -0.19$ for the values of parameters estimated here) which agrees quite well with the one observed in reality $\rho_{12}^{(obs)} = -0.24 \pm 0.16$ (95p.100 CI). Such a relationship partly avoids implementing more sophisticated models such as the bivariate Poisson regression model with three sets of parameters for additional modelling of covariance between H and A goals as proposed by Karlis and Ntzoufras (2003. This issue was investigated among others by Groll et al. (2018) on EUROs 2004–2012 who arrived at the same conclusion as ours but numerically. More details are given in appendix B on the analytical expression of this correlation allowing to test the conditions under which the simple Poisson loglinear model can be applied. Another approach allowing both positive and negative dependence was proposed by McHale and Scarf (2011) based on copula functions.

In such conditions, it is no big surprise that forecasts of the UEFA Champions League matches derived from our Poisson regression model are well calibrated and refined with good discrimination properties for both Home and Away Wins as shown in the summary table 5. Reliability amounts to 3.6 to 4.7 %, resolution to 25-30% and discrimination to 4-6%. Values are slight by better for Away than Home Wins with a trade-off between Home Win and under forecasted Draw with little discrimination for this last category. It looks as if this category stands apart from the two others. Bookmakers Odds Probability forecasts (ODD) remain superior to Poisson especially as far as Resolution (e.g., for Home Win 34.5 vs 29.5 % respectively) and Discrimination (7.7 vs 5.8%) are concerned thus indicating the excellence of their predictions as observed in many studies about association football competitions, this one being the first to our knowledge on the UEFA Champions League.

# Appendices

## A. The statistical model

Probability forecasting of WDL outcomes of matches can be carried out either directly or indirectly via the number of goals scored by the two teams. Score-line procedures has gained much popularity nowadays due to their multipurpose abilities for responding to the numerous types of soccer bets offered on the web (Simple; Draw no bet; Handicap; Correct score; Over/under; H…). That is the main reason why we choose this model as our central procedure to illustrate the probability scoring systems. Our model is a simple one based on a bivariate Poisson for the full-time score-lines

The Poisson regression model can be written for all teams $(i, j)$ involved in the matches $m(i, j) \in \mathcal{M}$ with score line $\left(y_{ij,1}^{(t)}; y_{ij,2}^{(t)}\right)$ at time (t) as:

$$y_{ij,k}^{(t)} \mid \lambda_{ij,k}^{(t)} \sim \mathcal{P}\left(\lambda_{ij,k}^{(t)}\right) \text{ for } k = 1,2$$

$$\log \lambda_{ij,1}^{(t)} = \eta + h + \beta_1 \Delta r_{ij}^{(t')} + \beta_2 \overline{r}_{ij}^{(t')}, \ \log \lambda_{ij,2}^{(t)} = \eta + \beta_1 \Delta r_{ji}^{(t')} + \beta_2 \overline{r}_{ij}^{(t')}$$

$$\Delta r_{ij} = r_i - r_j, \Delta r_{ji} = -\Delta r_{ij}, \ \overline{r}_{ij} = ½\left(r_i + r_j\right)$$

where $\eta$ is an intercept, $h$ is the home effect; $r_i^{(t)}$ the ELO ratings of team $i$ at time $t$ with $\Delta r_{ij}$ the difference in strength between the attacking team $i$ and defending team $j$, and $\overline{r}_{ij}$ representing their mean. For the convenience of computation and interpretation of the beta coefficients, the $r_i^{(t)}$'s are standardized as $r_i^{(t)} = \left(ELO_i^{(t)} - 1800\right)/150$ . Time $t'$ of ELO ratings is such that $t' < t$ and is chosen as close as possible to time of play $t$ of $m(i, j)$ e.g., on Monday for Tuesday C1 matches.

Independent prior distributions are set up on the parameters $\boldsymbol{\theta}$, i.e., $\eta \sim \mathcal{N}\left(\eta_0, \sigma_\eta^2\right)$, $h \sim \left(h_0, \sigma_h^2\right)$ and $\beta_k \sim_{ind} \mathcal{N}\left(b_k, \sigma_{\beta_k}^2\right)$ for $k = A, D$. Here, regarding the size of the training sample, we took them as non-informative priors with $\eta_0 = 0$, $\sigma_\eta^2 = 10^4$ ; $h_0 = 0$, $\sigma_h^2 = 10^4$ ; $b_k = 0$ and $\sigma_{\beta_k}^2 = 10^3$.

Table A-1: Parameter estimation of the Poisson Regression Model: Characteristics of the posterior distribution (ex from all samples 2014 to 2019 seasons)

| | mean | SD | Val 2.5 % | Val 97.5 % |
|---|---|---|---|---|
| Intercept | 0.117 | 0.034 | 0.049 | 0.183 |
| Home Effect | 0.329 | 0.043 | 0.246 | 0.413 |
| Team difference | 0.303 | 0.016 | 0.272 | 0.335 |
| Team average | 0.045 | 0.032 | -0.017 | 0.106 |

SD: standard deviation; 95 % credibility interval with low and up bounds (Val 2.5 %, Val 97.5%)

Knowing $\pi(\boldsymbol{\theta} \mid y)$ , the posterior distributions of $\boldsymbol{\theta}$, we can reconstruct the forecasting probabilities of elementary score lines of the future matches: $P_{ij}^{f}(u,v)=\Pr\left(Y_{ij,1}^{f}=u;Y_{ij,2}^{f}=v \mid \mathbf{y}\right)$ ($m$ in $m(i, j)$ skipped to reduce the burden of notation) as the mean of the posterior distribution of the probability $\Pr\left(Y_{ij,1}^{f}=u;Y_{ij,2}^{f}=v\right)$ i.e.,

$$\begin{aligned}\Pr_{post}\left(Y_{ij,1}^{f}=u;Y_{ij,2}^{f}=v\right)&=\int \Pr\left(Y_{ij,1}^{f}=u;Y_{ij,2}^{f}=v \mid \boldsymbol{\theta}\right)\pi(\boldsymbol{\theta} \mid y)\,d\boldsymbol{\theta}\\ &=\int \Pr\left(Y_{ij,1}^{f}=u \mid \lambda_{ij,1}^{f}\right)\Pr\left(Y_{ij,2}^{f}=v \mid \lambda_{ij,2}^{f}\right)\pi\left(\lambda_{ij,1}^{f},\lambda_{ij,2}^{f} \mid y\right)d\lambda_{ij,1}^{f}d\lambda_{ij,2}^{f}\end{aligned}$$

Due the assumption of conditional independence of the $y_{m(ij),k}^{(t)}$, this reduces to the posterior expectation of the product of the marginal probabilities of the Home and Away scores:

$$P_{ij}^{f}(u,v)=\mathrm{E}\left(\lambda_{ij,1}^{f}\lambda_{ij,2}^{f}\exp\left[-(\lambda_{ij,1}^{f}+\lambda_{ij,2}^{f})\right]/u!v! \mid \mathbf{y}\right),$$

where $\mathbf{y}$ includes ex-ante scorelines of the 3 previous seasons plus ELO9 ratings prior to day of play.

Eventually, the probabilities of $X_{ij}$ outcomes between teams $i$ and $j$ in terms of WDL are obtained as sums of the previous elementary score-line probabilities,

$$\Pr\left(X_{ij}=[1] \mid \mathbf{y}\right)=\sum\nolimits_{u,(v<u)}P_{ij}^{f}(u,v \mid \mathbf{y}),\ \Pr\left(X_{ij}=[2] \mid \mathbf{y}\right)=\sum\nolimits_{u}P_{ij}^{f}(u,u \mid \mathbf{y}),$$

the last category being obtained by the complement to 1 of the sum of the two previous ones.

## B. Correlation between Home (H) and Away (A) goals induced by the Poisson model

For the sake of simplicity and knowing that the little contribution of ELO rating average (Table B-1, and P value=0.157 for a deviance of 2 with one degree of freedom), we consider the Poisson loglinear model under its reduced form with just an intercept and a covariate difference between rates of teams, with the following condensed notations. Let $\mathbf{y}_{m}=\left(y_{m,1},y_{m,2}\right)$ be the score-lines of match $m$ between teams 1 and 2 referring to H and A goals scored by these two teams respectively. The model is written as:

1) $\mathbf{y}_{m,k} \mid \lambda_{m,k} \sim P(\lambda_{m,k})$ for $k = 1,2$

2) $\log \lambda_{m,1} = \eta_1 + \beta \Delta r_m$, $\log \lambda_{m,2} = \eta_2 - \beta \Delta r_m$,

where $\eta_1, \eta_2$ are intercepts for H and A goals respectively and $\beta$ is the regression coefficient on $\Delta r_m = r_1 - r_2$ team rates (e.g., ELO).

Classically, in regression GLM model, inference about the parameters, here $\boldsymbol{\theta} = (\eta_1, \eta_2, \beta)$ is based on Maximum Likelihood procedures by conditioning on observed values of the $\Delta r_m \sim_{iid} \mathrm{N}(0, \gamma)$'s random variables (Pawitan, 2013, page 278). However, as $\Delta r_m$ pops into both H and A parameters terms, one may expect some correlation between H and A goals scored in a match, here negative due to opposite signs. The question arises on whether this correlation generated by sampling the $\mathbf{y}_{m,k}$'s in a population of matches can be expressed in a closed form.

To do that, we will rely on basic results on the exponential transformation $Y = \exp(X)$ of a normal distribution $X \sim \mathcal{N}(\mu_X, \sigma_X^2)$. These are:

$$\mathrm{E}(Y) = \mu_Y = \exp(\mu_x)\left[\exp(\sigma_X^2 / 2)\right];$$

$$\mathrm{Var}(Y) = \sigma_Y^2 = \mu_Y^2\left[\exp(\sigma_X^2) - 1\right]; \ \mathrm{Cov}(Y_1, Y_2) = \sigma_{Y_1Y_2} = \mu_{Y_1}\mu_{Y_2}\left[\exp(\sigma_{X_1X_2}) - 1\right].$$

Now, the law of total variance says upon conditioning and deconditioning for an auxiliary variable: $\mathrm{Var}(Y) = \mathrm{E}_Z\left[\mathrm{Var}(Y \mid Z = z)\right] + \mathrm{Var}_Z\left[\mathrm{E}(Y \mid Z = z)\right]$. In our case, $Y$ being Poisson and $Z$, its parameter $\lambda$, one has $\mathrm{E}(Y \mid \lambda) = \mathrm{Var}(Y \mid \lambda) = \lambda$, and $\mathrm{E}(\lambda)$ as $\mathrm{Var}(\lambda)$ follow from $\log(\lambda_k) \sim \mathcal{N}(\eta_k, \gamma)$ so that:

$$\mathrm{Var}(y_{m,k}) = \mu_k\left\{1 + \mu_k\left[\exp(\beta^2\gamma) - 1\right]\right\}.$$

Similarly, assuming conditional independence between $y_{m,1}$ ,and $y_{m,2}$ one gets for the covariance:

$$\mathrm{Cov}(y_{m,1}, y_{m,2}) = \mu_1\mu_2\left[\exp(-\beta^2\gamma) - 1\right],$$

where $\mu_k = \exp(\eta_k)\left[\exp\left(\beta^2\gamma/2\right)\right]$, $k=1,2$], and the correlation $\rho_{12}$ between $y_{m,1}$ and $y_{m,2}$ is :

$$\rho_{12} = \frac{\sqrt{\mu_1\mu_2}\left[\exp\left(-\beta^2\gamma\right)-1\right]}{\sqrt{1+\mu_1\left[\exp\left(\beta^2\gamma\right)-1\right]}\sqrt{1+\mu_2\left[\exp\left(\beta^2\gamma\right)-1\right]}}.$$

In this model $\beta^2\gamma \geq 0$ and $\rho_{12}$ is always negative for $\beta \neq 0$. Using the values of parameters estimated on the 2014-2019 sample (6x124=744 matches) i.e., $\eta_1 = 0.1197$, $\eta_2 = 0.1197+0.3293=0.4490$, $\beta = 0.3014$ and $\gamma = 2$ (due to standardization of team rates $r_k$), one obtains $\rho_{12} = -0.1894$.

This is in good agreement with the observed value of this correlation coefficient $\rho_{12}^0 = -235$ and 95% CI (-0.076,0314). Values of this correlation (free of any model) show the same constant negative tendency over periods in the range of -0.101 (2017) to -0.298 (2014).

Table B-2: Estimation of the correlation coefficient between Home and Away goals scored in a match and its corresponding 95% confidence interval (Low, Up)

| season | 2014 | 2015 | 2016 | 2017 | 2018 | 2019 | 2014-19* |
|---|---|---|---|---|---|---|---|
| Low | -0.450 | -0.375 | -0.313 | -0.272 | -0.528 | -0.406 | -0.394 |
| Est | -0.298 | -0.213 | -0.145 | -0.101 | -0.389 | -0.248 | -0.235 |
| Up | -0.128 | -0.039 | 0.031 | 0.076 | -0.229 | -0.075 | -0.062 |

*Pooled within-season

Similar results were observed using different transformation of H and A goals e.g., Ascombe, Tukey and log as well as non-parametric estimators of correlation: Spearman and Kendall.

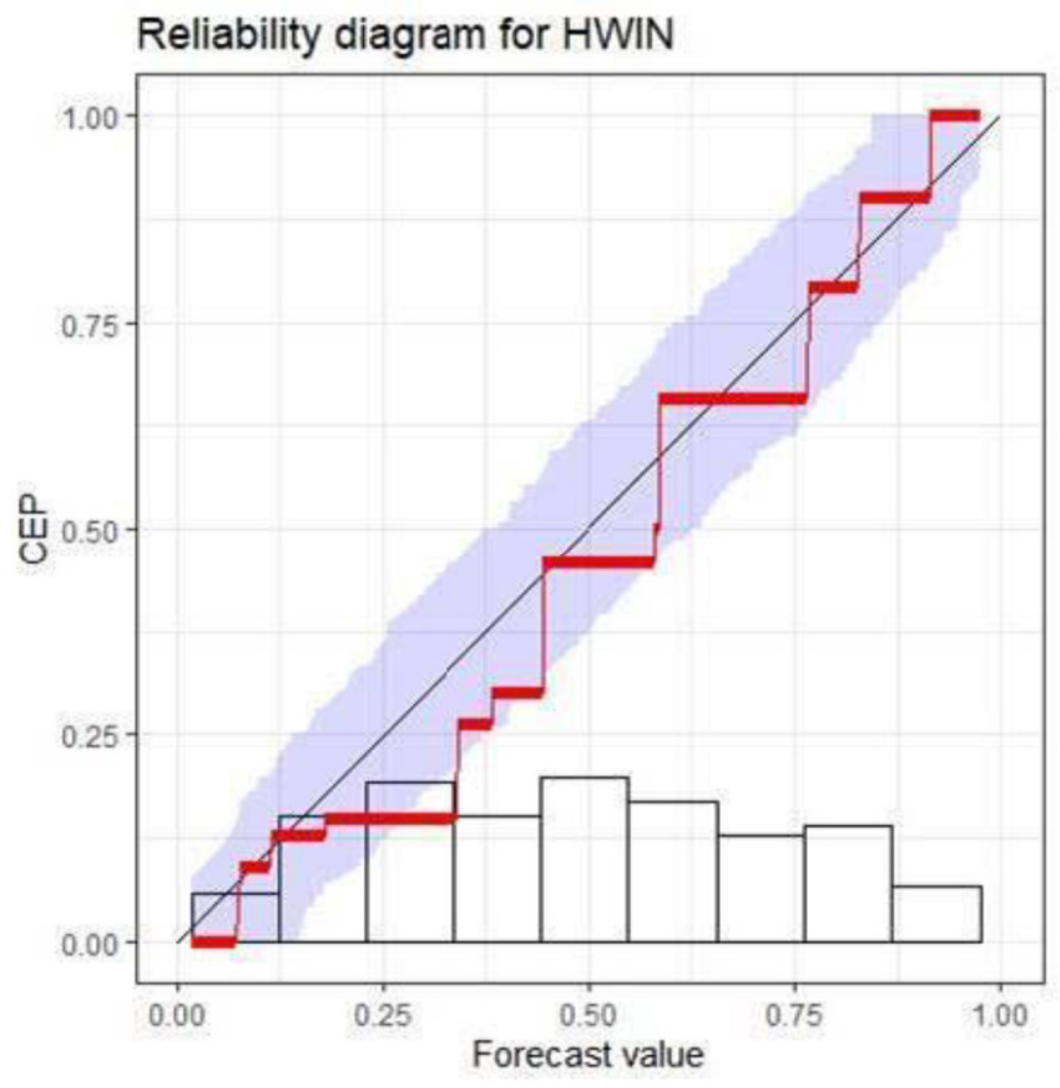

1-HomeWin Poisson Iso-Reg

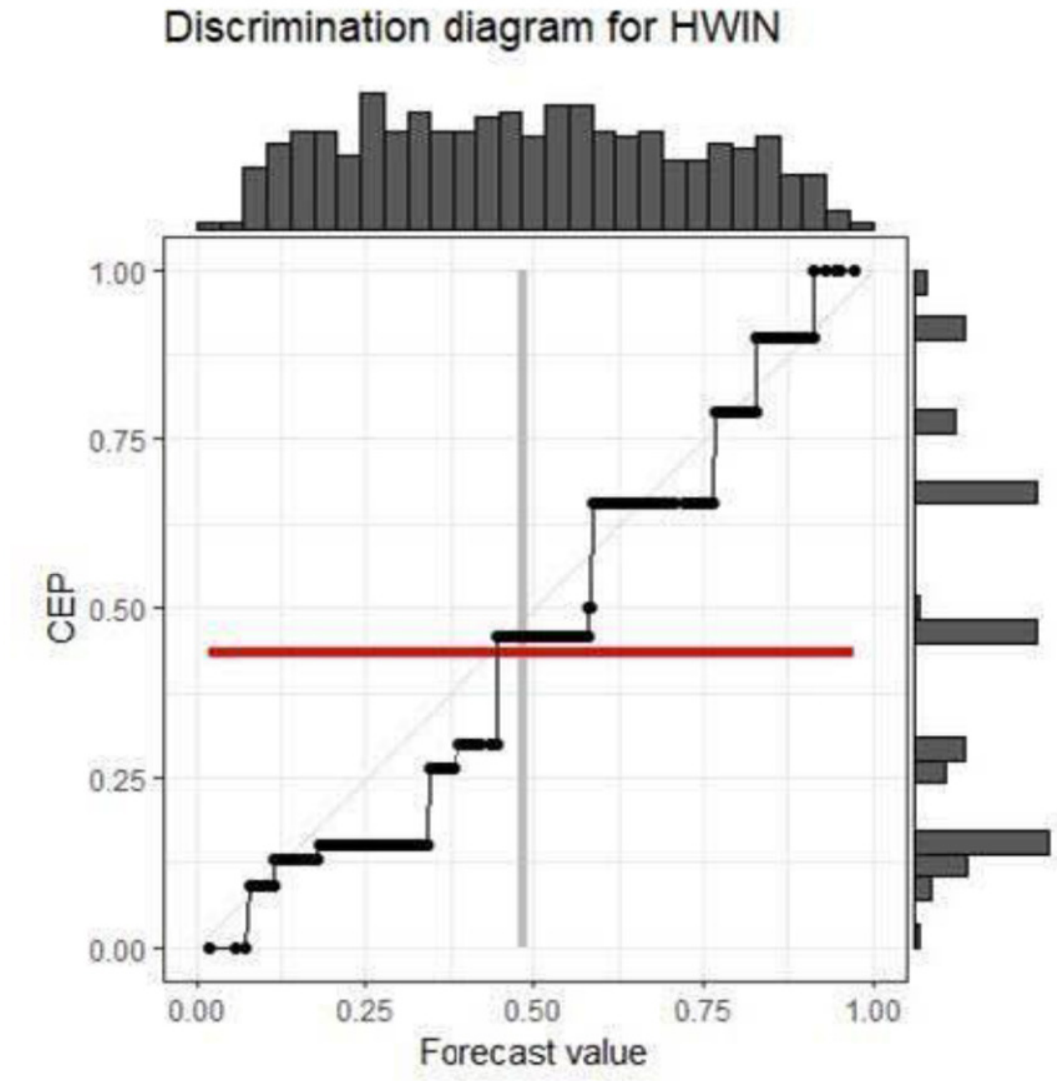

3-HomeWin Poisson Iso-Reg

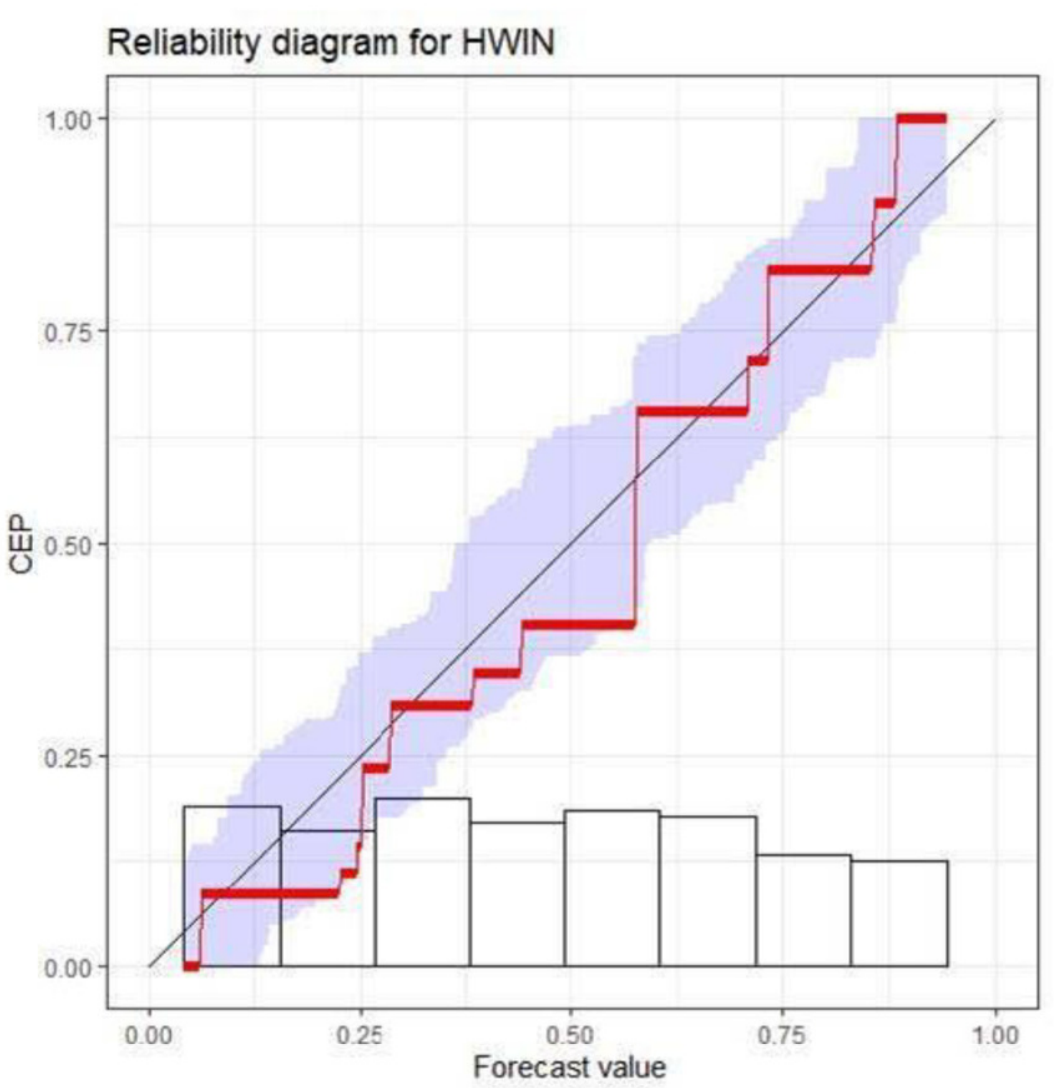

2-HomeWin Odd Iso-Reg

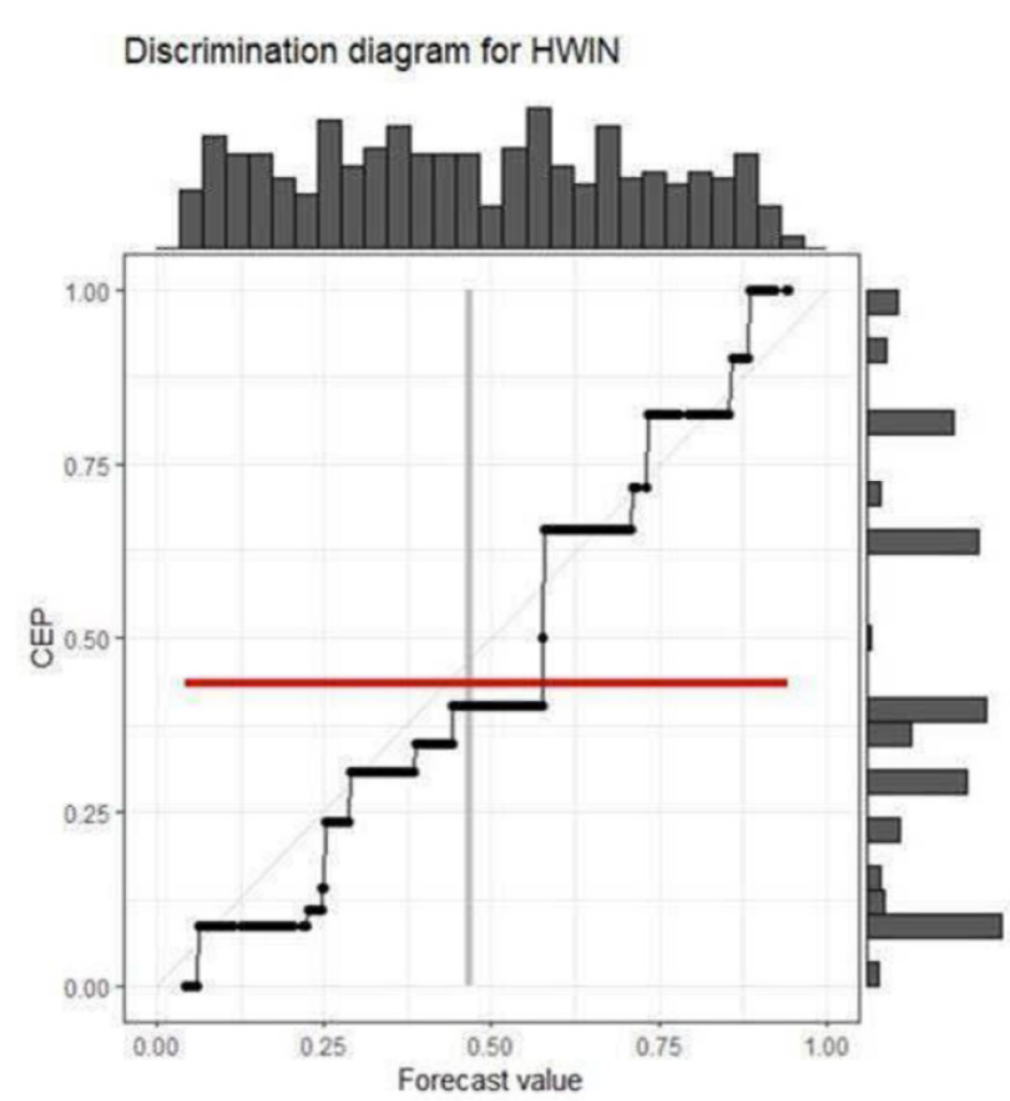

4-HomeWin Odds Iso-Reg

Figure 1: Reliability Diagrams for Home Win Probability Forecasts with plots of the Conditional Probability Events (CEP) against the Forecast Probability Values via Iso Regression

In 1 and 2: Reliability with Point 95% Consistency Bands; 3 and 4: Discrimination with histograms of the marginal distribution of the original forecast values (top) and at right, that of the calibrated probability values.

For both methods predictions, there is some evidence of over-forecasting at forecasted values below 0.6

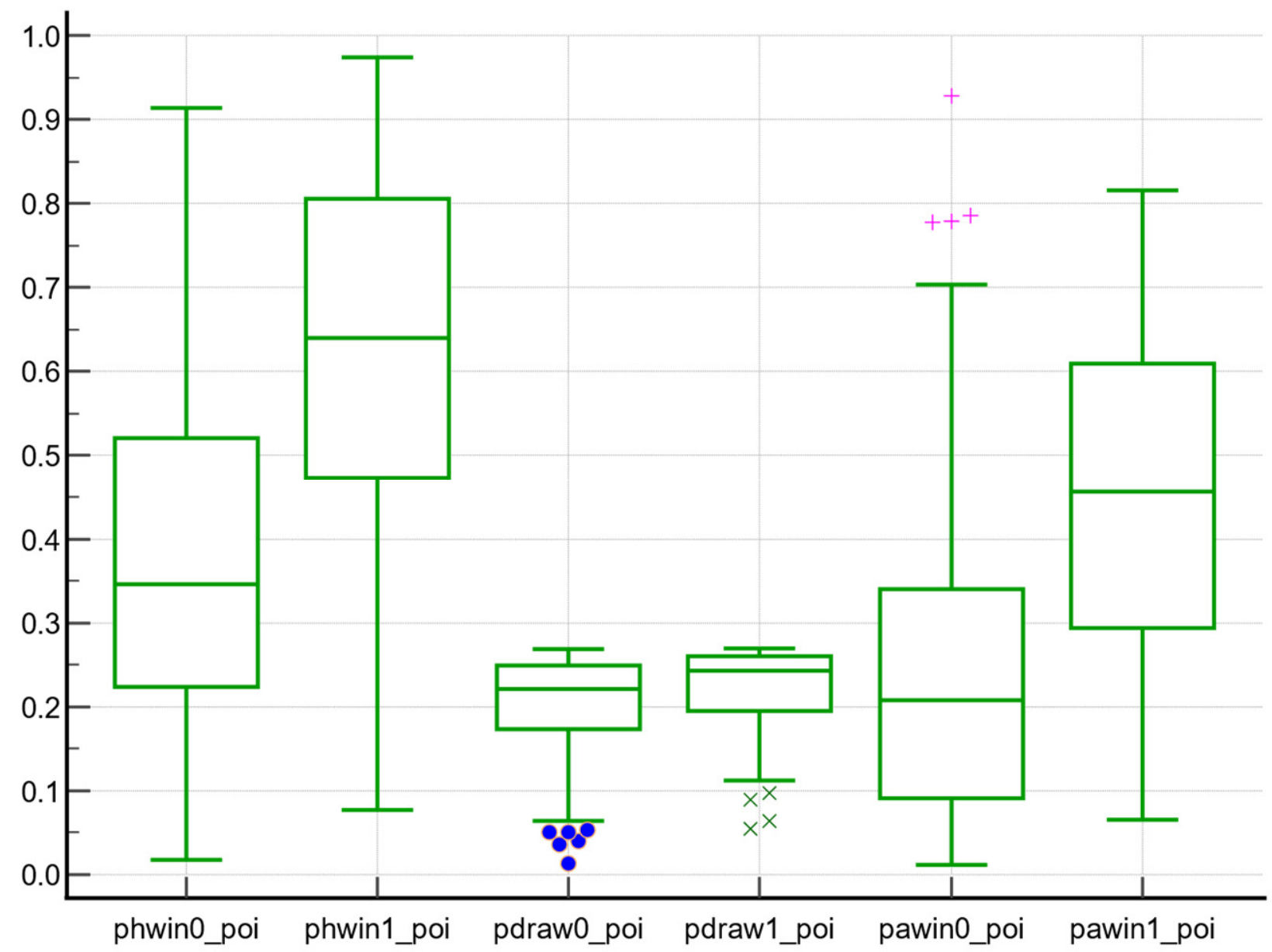

1. *Home Win, Draw and Away Win for the Poisson Model (POI)*

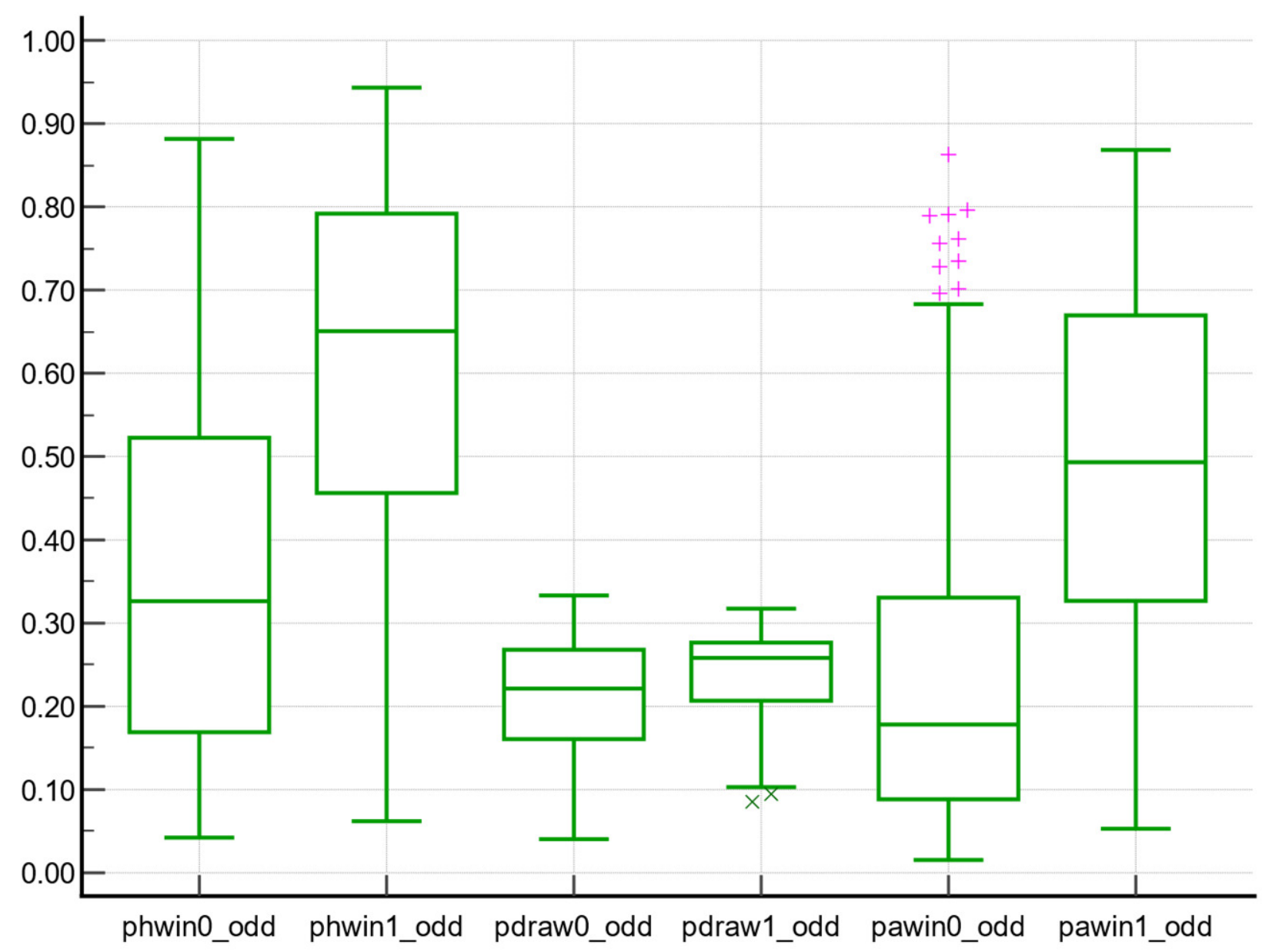

2. *Home Win, Draw and Away Win for the Bookmakers Odds Probabilities (ODD)*

*Figure 2: Box-plots of the conditional distributions of probabilistic forecasts of binary events (HWIN, DRAW and AWIN) given the observed outcomes (X=0) and (X=1) and according to the forecasting procedures: 1) Poisson loglinear regression (POI) and 2) Bookmakers Odds Probabilities (ODD)*